\documentclass{aa}
\usepackage{subfig}
\usepackage{graphicx}
\usepackage{tabularx}  
\usepackage{txfonts}

\usepackage{hyperref} 
\hypersetup{colorlinks=true,linkcolor=blue,citecolor=blue,filecolor=blue,urlcolor=blue,}
\makeatletter
\renewcommand*\aa@pageof{, page \thepage{} of~
\pageref*{LastPage}}
\makeatother

\newcommand{\kms}{$\textrm{km/s}\:$}
\newcommand{\kpvsys}{$K_{p}V_{sys}\ $}

\begin{document}


\title{Small but mighty: High-resolution spectroscopy of ultra-hot Jupiter atmospheres with compact telescopes}

\subtitle{Transmission spectrum of KELT-9\,b  with Wendelstein's FOCES spectrograph}

\author{N. W. Borsato\inst{1,4}
          \and
            H. J. Hoeijmakers\inst{1}
          \and
            D. Cont\inst{2}
          \and
            D. Kitzmann\inst{3}
          \and
            J. Ehrhardt\inst{2}
          \and
            C. G\"{o}ssl\inst{2}
          \and
            C. Ries\inst{2}
            \and
            B. Prinoth\inst{1}
          \and
            K. Molaverdikhani\inst{2}
          \and
            B. Ercolano\inst{2}
          \and
            H. Kellerman\inst{2}
          \and
            Kevin Heng\inst{5,6}
}

\institute{Lund Observatory, Division of Astronomy and Theoretical Physics, Lund University, Box 43, 221 00 Lund, Sweden\\
\email{n.borstato@astro.lu.se}
\and
Universit\"ats-Sternwarte M\"unchen, Fakult\"at f\"ur Physik, Ludwig-Maximilians-Universit\"at M\"unchen, Scheinerstrasse 1, 81679 M\"unchen, Germany
\and
University of Bern, Center for Space and Habitability, Gesellschaftsstrasse 6, CH-3012, Bern, Switzerland
\and
Research Centre for Astronomy, Astrophysics and Astrophotonics, Macquarie University, Balaclava Road, Sydney, NSW 2109 Australia
\and
ARTORG Center for Biomedical Engineering Research, Murtenstrasse 50, CH-3008, Bern, Switzerland
\and
Astronomy \& Astrophysics Group, Department of Physics, University of Warwick, Coventry CV4 7AL, United Kingdom
}

\date{Received August 9, 2023; accepted November 20, 2023}

\abstract{When observing transmission spectra produced by the atmospheres of ultra-hot Jupiters (UHJs), large telescopes are typically the instrument of choice given  the very weak signal of the planet's atmopshere. The aim of the present study is to demonstrate that, for favourable targets, smaller telescopes are fully capable of conducting high-resolution cross-correlation spectroscopy. We apply the cross-correlation technique to data from the 2.1m telescope at the Wendelstein Observatory, using its high-resolution spectrograph FOCES, in order to demonstrate its efficacy in resolving the atmosphere of the UHJ KELT-9\,b. Using three nights of observations with the FOCES spectrograph and one with the HARPS-N spectrograph, we conduct a performance comparison between FOCES and HARPS-N. This comparison considers both single-transit and combined observations over the three nights. We then consider the potential of 2m class telescopes by generalising our results to create a transit emulator capable of evaluating the potential of telescopes of this size. With FOCES, we detected seven species in the atmosphere of  KELT-9\,b: Ti\,II, Fe\,I, Fe\,II, Na\,I, Mg\,I, Na\,II, Cr\,II, and Sc\,II. Although HARPS-N surpasses FOCES in performance thanks to the mirror of the  TNG, our results reveal that smaller telescope classes are capable of resolving the atmospheres of UHJs given sufficient observing time. This broadens the potential scope of such studies, demonstrating that smaller telescopes can be used to investigate phenomena such as temporal variations in atmospheric signals and the atmospheric loss characteristics of these close-in planets.
}

\keywords{Planets and satellites: atmospheres --
Planets and satellites: gaseous planets --
Techniques: spectroscopic
}

\maketitle
\section{Introduction}
\label{sec:introduction}
In ground-based high-resolution spectroscopic transit observations of exoplanet atmospheres, telescope size often determines the scale of the scientific aims, with larger telescopes typically favoured for these types of observations. Atmospheric observations of exoplanets suffer from photon scarcity, and as such, larger mirrors prove indispensable, collecting more photons and compensating for the feeble planet signal, which can be as low as a few parts per million relative to the flux of the host star in the case of hot Jupiters, and even lower for smaller planets~\citep{Brogi_2012}. This scarcity often results in the subtle absorption of the exoplanet signal becoming indistinguishable within the noise of the observation. These challenges limit detailed observational studies to certain subclasses of exoplanets, which include the hottest exoplanets known to date, ultra-hot Jupiters (UHJs)~\citep{Parmentier_2018}.

UHJs orbit close to their host star and serve as excellent targets for atmospheric studies thanks to their typically inflated atmospheres, which result in large transit depths~\citep{Baraffe_2010, Merritt_2021}. The extreme irradiation these planets undergo as a result of their proximity to their host star leads to unique atmospheric effects, such as molecular dissociation~\citep{Kitzmann_2018,Arcangeli_2018,Parmentier_2018}, cloud-free day sides~\citep{Helling_2021}, equatorial jets moving at velocities of kilometres per second~\citep{Showman_2011}, and mass-loss through photo-evaporation~\citep{Lowson_2023}.

Owing to their unique atmospheric properties, UHJs serve as laboratories for exoplanet science. Their atmospheres are generally simpler to interpret due to relatively stable equilibrium conditions at pressures of higher than $10^{-4}$ bars~\citep{Kitzmann_2018,Fossati_2021,Lee_2022}. Though non-local thermodynamic equilibrium (NLTE) effects and deviations from chemical equilibrium can occur in the upper and lower layers, respectively~\citep{Fossati_2021, Arcangeli_2021}, the majority of the transmission spectra of UHJs are predicted to probe regions where equilibrium assumptions hold~\citep{Kitzmann_2018,Lee_2022}. Indeed, they are ideal test targets for exploring the extent to which telescope size can limit atmospheric studies. Small telescopes are the workhorses of astronomy, enabling vital research that may be considered `high risk' or may only yield results in the long-term. The lower subscription factors of these classes of telescope also enable the possibility of multiple observations of the same target, which may produce scientific results of the same quality as larger telescopes when combining these observations, and would also enable studies of the time evolution of the atmospheres of these exoplanets.

There is a growing list of telescopes that have successfully conducted high-resolution spectrographic observations of UHJs, including the Nordic Optical Telescope~\citep[with a mirror diameter of 2.5m][]{Bello_Arufe_2022}, as well as the Fred Lawrence Whipple Observatory~\citep{Lowson_2023}, both of which have produced detections within a single transit. In the present work, we aim to add another 2m class instrument to the pool of telescopes able to produce an atmospheric detection of a UHJ atmosphere in a single transit, namely the FOCES spectrograph on the 2m Wendelstein telescope located in the Bavarian alps~\citep{Pfeiffer_1998}. Additionally, we aim to expand the discussion to consider the full potential of 2m telescopes in general, and specifically in observing known UHJs. The target of choice is KELT-9\,b, an UHJ with an equilibrium temperature of approximately 4000\,K that has proven to be a highly observable target for benchmark cases such as that presented here~\citep[e.g.][]{Hoeijmakers_2018,Yan_2019,Bello_Arufe_2022,Lowson_2023}.

Furthermore, we aim to compare the performance of FOCES to that of the HARPS-North (HARPS-N) high-resolution spectrograph on the Telescopio Nazionale Galileo (TNG), which is an instrument used as standard in the study of exoplanet atmospheres~\citep[e.g.][]{Casasayas_Barris_2019,Stangret_2020,Stangret_2021,Cont_2021,Borsa_2022,Cont_2022,Pino_2022}. Following this comparison, we generalise and extrapolate our results to comment on the potential of 2m class telescopes and their ability to resolve exoplanet atmospheres.

\section{Observations}
We used FOCES to observe three transits of KELT-9\,b on 18$^{}$ July 2022, 2$^{}$ September 2022, and 5$^{}$ September 2022 (only a partial transit). Each transit was observed in time series, taking continual exposures of 300\,s in length throughout the transit, plus some additional exposures out of transit to obtain a stellar baseline without the planet signal. The other data set used is a single transit observation taken using the HARPS-N spectrograph on the TNG, which is available in a public data archive and was initially published in~\cite{Hoeijmakers_2018}. The transit data were obtained on 31$^{}$ July 2017, with an integration time of 600s per exposure frame. The observation encompassed the entire transit and baseline measurements. Table~\ref{tab:ObservationalSummary} provides a summary of the observations taken with each spectrograph, including the relevant characteristics of the instruments and the relative amount of stellar baseline taken compared to the length of the transit. Air-mass plots for each transit are shown in Fig. \ref{fig:airmassplots}; these depict where on the sky the transit occurs and during what epoch of the night.

\begin{figure}[ht]
    \centering
    \includegraphics[width=\linewidth]{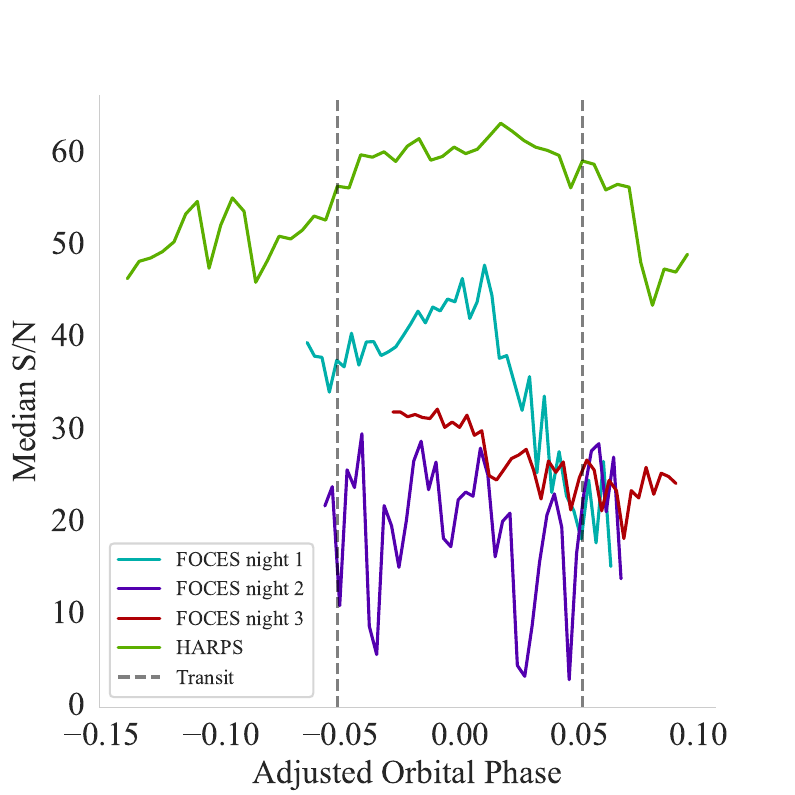}
    \caption{Time trend of the median S/N for each transit observation of KELT-9\,b as a function of orbital phase. The HARPS-N S/N values have been scaled to values that would be expected to be produced for the exposure times taken by the FOCES spectrograph. The HARPS-N transit covers the transit of the entire planet, and a substantial amount of the baseline. Additionally, the S/N is consistently higher throughout the entire observation compared to the observations taken by FOCES. Signal quality also drops off for all the FOCES nights towards the end of each night.}
    \label{fig:snr_plot}
\end{figure}

Using the approach described by~\cite{Borsato_2023}, we measured the average quality of the data by taking the median signal-to-noise ratio (S/N) over all the echelle orders; this allows us to track the progression of the data quality over time throughout the transit. The progression of the
S/N over time for each observation is shown in Fig. \ref{fig:snr_plot} as a function of orbital phase, with the beginning and end of the transit marked. The figure provides a good overall picture of how the observation progressed throughout each night. So as to be able to compare the observations, we divided the S/N values for the HARPS-N night by a factor of $\sqrt{2}$ (assuming a Poisson noise distribution) in order to correct for the time difference in the exposures.

Among the three FOCES nights, the first exhibited the highest S/N, which later dropped by a factor of two towards the end of the observation. The night began with an overcast sky of cirrus clouds. However, these clouds began to dissipate just before observations began. Approximately midway through observations, the signal quality dropped due to a slow build-up of clouds towards the end of the night. The S/N on the second night of observations showed significant variation throughout the entire transit. The entire night was characterised by the presence of cirrus clouds of varying opacity. Nevertheless, these clouds did not completely obscure the sky, allowing some measurements to be obtained. Throughout the night, the signal dropped steadily and oscillated periodically. The third night was mainly clear, with occasional thin and narrow cirrus clouds passing by, but due to twilight timing constraints, only a partial transit could be obtained. As the night progressed, the signal dropped slightly during the exposure times, likely due to fast-moving clouds partially obscuring the observations. However, these clouds were transient, leading to a patchwork drop in night flux rather than a consistent decrease. The HARPS-N spectrograph outperforms all three FOCES observations by a factor of approximately two in terms of S/N, and no significant problems with the weather were noted.

\section{Data processing}
FOCES spans a wavelength range of between 383 and 885\,nm at a spectral resolution of \textit{R} $\sim$ 70,000~\citep{FOCES_1998,Brucalassi_et_2012,Wang_et_al_2017}. The primary science products derived from the spectrograph's pipeline include two-dimensional echelle data, recorded errors, and blaze profiles. We stitched the overlapping pixels of the echelle orders by averaging the overlapping flux values and using the recorded errors as weights, thereby generating a one-dimensional spectrum of each exposure. The HARPS-N spectrograph spans a wavelength range from 387 to 690\,nm with a resolution of \textit{R} $\sim$ 115,000. The scientific products available from these observations include the two-dimensional echelle orders and the blaze-corrected, stitched, and resampled one-dimensional spectra.

The gathered data necessitated the removal of systematic signals extraneous to the planet. We adopted the methodology outlined in~\cite{Hoeijmakers_2020b} for this process, which entailed the elimination of atmospheric telluric contamination, velocity offsets, the interstellar medium (ISM), the host star signal, and residual contamination in the continuum. The same process for the removal of systematic signals was uniformly applied to both the FOCES and HARPS-N datasets, despite their differing S/N levels. To remove the telluric lines produced by water and carbon dioxide, we used the one-dimensional spectra. We were able to model the telluric lines using the software package \texttt{molecfit}~\citep{Smette_2015,Kausch_2015}, thereby creating an atmospheric telluric model across the full wavelength range of the data analysed. After modelling, we remove the telluric lines from the observation by dividing by a telluric model produced for each exposure.

Furthermore, there are two velocity offsets that require correction; these are caused by two Doppler shifts, those arising from the Earth--Sun barycentric velocity and the radial velocity of the star produced by the planet’s gravitational pull. The correction leaves the spectra at a constant radial velocity shift set by the systemic velocity of approximately \text{-}18 \kms \citep{Gaudi_2017}. The barycentric velocity correction is calculated using the Python library \texttt{astropy}, while the radial velocity offset is available from the discovery paper~\citep{Gaudi_2017}.

To remove the signal of the host star, we compute the average spectrum over all the exposures of KELT-9\,b, {and divide each exposure} by this average; this removes the static components of the spectrum while maintaining the time transient properties, such as the signal from the planet. Dividing by the average produces a residual continuum shape, which needs to be flattened using a median filter.

Finally, any bad pixels remaining are flagged as not a number (NaN) using a running standard deviation, and any values exceeding 5$\sigma$ are flagged. The data are further cleaned by applying a Gaussian high-pass filter with a bandwidth of 80\,\kms, and once again marking any outliers beyond the 5$\sigma$ limit. Finally, any residual ISM effects are manually flagged as NaNs. The NaNs generated from this process create an outlier mask that enables us to bypass problematic pixels during the cross-correlation process. All these steps are applied through the cross-correlation software package \texttt{tayph}\footnote{https://github.com/Hoeijmakers/tayph}{, which is under development at Lund Observatory and forms the basis for similar recent cross-correlation studies \citep[e.g.][]{Hoeijmakers_2020_WASP121b,Hoeijmakers_2023,Prinoth_2022,Borsato_2023,Prinoth_2023}.}

\begin{table*}
    \centering
    \caption{Summary information of the transit observations used in this study}
    \label{tab:ObservationalSummary}
    \begin{tabularx}{\textwidth}{lXXXXXXX}
    \hline \hline
        Date & Spectrograph & Wavelength Range [nm] & Resolution & S/N [300s] & Exp-Time [s] & Exp-Number & Baseline Fraction \% \\ \hline
        31-07-2017 & HARPS-N & 387--690 & 115,000 & 56 & 600 & 49 & 57\\
        18-07-2022 & FOCES & 380--885 & 70,000 & 34 & 300 & 41 & 18 \\
        02-09-2022 & FOCES & 380--885 & 70,000 & 26 & 300 & 41 & 33 \\
        05-09-2022 & FOCES & 380--885 & 70,000 & 30 & 300 & 39 & 17 \\
    \hline
    \end{tabularx}
    \tablefoot{This table summarises the details of the four observations used in this study. Columns denote the date of observation, the spectrograph used, the operational wavelength range in nanometres, the spectral resolution, the S/N ratio per 300 seconds of exposure time, with the HARPS-N scaled through Poisson noise assumptions, the exposure time in seconds, the total number of exposures, and the percentage of observations outside of transit.}
\end{table*}

\section{The cross-correlation technique}
\label{sec:ccfapproach}

To search for atmospheric signals, we applied the cross-correlation method. This innovative approach was first introduced by \citet{Snellen_2010_CC}, enabling the detection of CO in the transmission spectrum of HD 209458 b. The essence of this method is to combine the flux contributions of specific spectral lines within a planet's transmission spectrum, thereby effectively generating a weighted average of the spectral line profile for a given species based on its known line positions.

However, the variable radial velocity of the planet introduces a shift in the spectral line positions within the wavelength space. To accommodate this variability, the technique employs Doppler shifting to adjust the known line positions to different radial velocity values. This step allows the spectrum to be scanned continuously for these shifted lines.

The real strength of this method becomes evident when the predicted line positions align with the planet's absorption spectrum. This alignment leads to a correlation peak, which signifies the successful detection of a species within the planet's atmosphere.

In order to carry out this method, it is necessary to construct a spectral template, $T$. This template calculates the transmission spectrum of either the entire planet or an individual species. It is then used to generate a set of weights, which are assigned to the anticipated line positions of the species under investigation:

\begin{equation}
\hat{T}_i(v) = \frac{T_i(v)}{\Sigma T_i(v)}
.\end{equation}

Here, the template function, $T_i(v)$, is dependent on the radial velocity, where $i$ refers to the
species under investigation, and $v$ denotes the radial velocity of the template. The weights are scaled so that the total sum of the templates is equal to one, with large weight values allocated to the line positions. This weighted average is then employed to calculate the mean line depth, as illustrated in the following equation:

\begin{equation}
c(v) = \sum_{i=0}^{N} x_i \hat{T}_i(v)
\label{eq:ccf}
,\end{equation}

where $c(v)$ represents the computed mean depth of absorption lines, where each line is weighted relative to its position at a specific Doppler shift. The dependence on the velocity term implies that the final value obtained from the sum will change depending on the Doppler shift applied to the calculation. The total quantity of evaluated wavelength bins is represented by $N$. Each $x_i$ corresponds to the spectral data that align with the same wavelength segment $i$ as the reference template.

\begin{figure}[ht] 
    \centering 
    \includegraphics[width=\columnwidth]{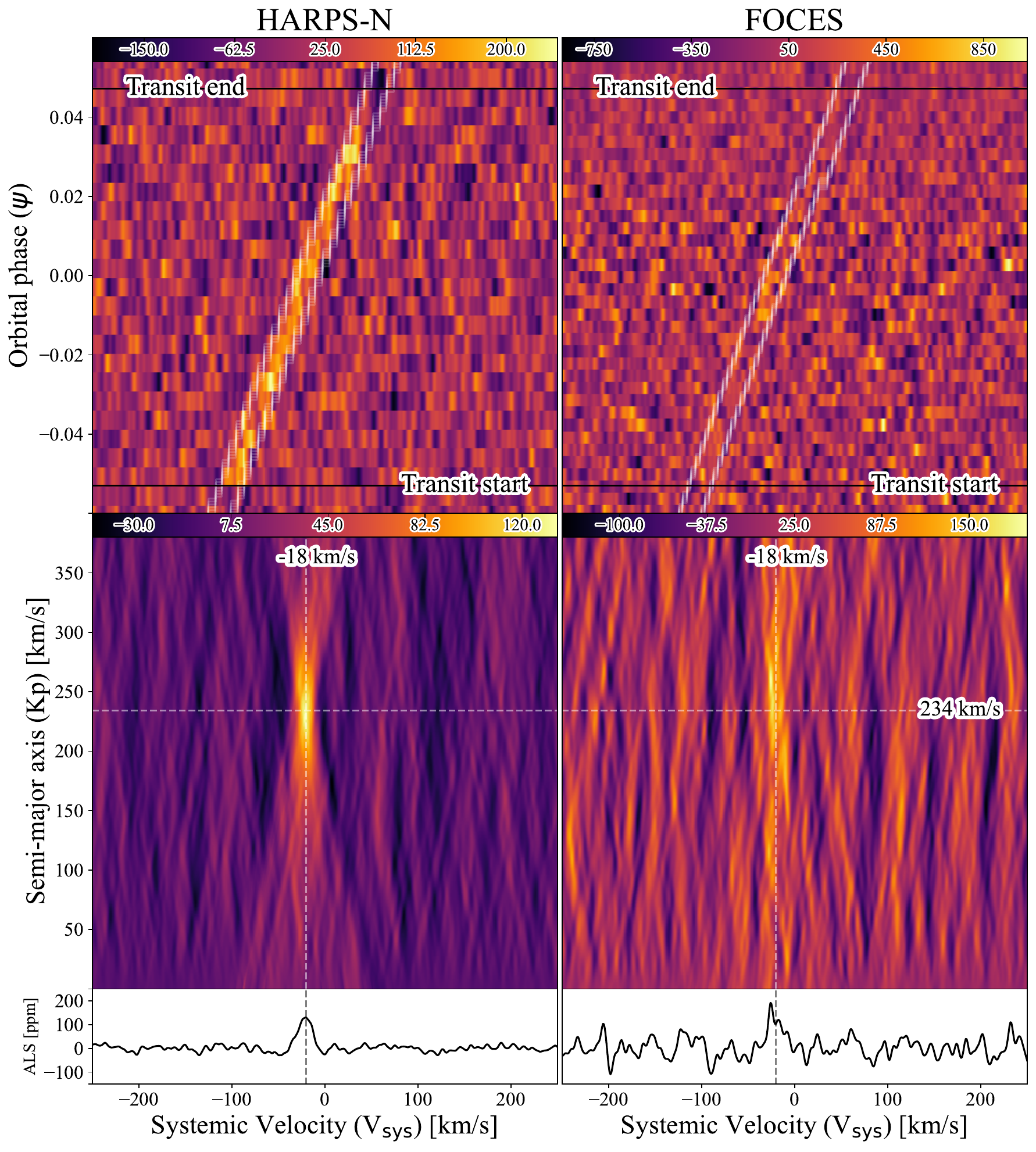} 
    \caption{Cross-correlation results for searching for Fe\,I in a single transit. On the left are the results for the HARPS-N spectrograph on the TNG, and on the right we show the results for the FOCES spectrograph for the night of 18$^{}$ July 2022. \textit{Top Panel:} Cross-correlation maps for Fe\,I. The pixels covering the trace of the planet are outlined to help guide the eye to where to expect the signal. \textit{Middle panel:} \kpvsys\ maps for each Fe\,I detection. The detections are the bright features in the centres of the plots. The dashed cross-hairs show the row where the one-dimensional CCF (bottom panel) has been extracted and the expected systemic velocity of KELT-9\,b (-20 \kms). \textit{Bottom Panels:} One-dimensional CCFs for the single transits, showing the average line depth for the given signal. The dashed line shows the expected systemic velocity of KELT-9\,b, which is the same as in \kpvsys\ map. }
    \label{fig:cross_correlation_results_one_night} 
\end{figure}

The cross-correlation function (CCF) leverages the fact that the radial velocity of a planet changes throughout its transit, leading to a shift in the line positions of the transmission spectrum. By applying Eq. \ref{eq:ccf}, a peak will form at the correct velocity position if the species is present in the planet's atmosphere. This process results in an absorption trace on a two-dimensional grid of time versus planetary radial velocity, as shown in the top panel of Fig. \ref{fig:cross_correlation_results_one_night}.

After applying the CCF, further processing steps are necessary to detect an atmospheric signal. The process consists of several sequential actions. Initially, residual broadband variations stemming from the removal of the stellar signal are eliminated. Subsequently, the Rossiter-McLaughlin (RM) effect~\citep{Rossiter_RM, McLaughlin_RM}, a phenomenon caused by the planet temporarily distorting the line-of-sight velocity of the star by obscuring part of the surface of the rotating star (commonly known as the Doppler shadow) is modelled and removed. The methodology adopted for
removal of broadband variation and compensation for the Rossiter-McLaughlin effect follows that of~\cite{Hoeijmakers_2020_WASP121b}, who successfully conducted a chemical inventory of WASP-121 b. Finally, we remove any residual stellar aliases by applying the same vertical detrending adapted from~\cite{Prinoth_2022} and implemented by~\cite{Borsato_2023} to remove residual cross-correlation signatures coming directly from the star's spectra.

It is also possible to enhance the S/N of a cross-correlation function by co-adding all pixels along the trace of the planet. This is achieved by shifting all exposures into the time-independent rest frame of the planet using its orbital phase and inclination. This process can be executed with the following equation:

\begin{equation}
v_{rv} = v_{\textrm{orb}}\sin{2\pi\phi}\sin{i} + v_{\textrm{sys}}
\label{eq:orbital_equation}
,\end{equation}

where $v_{orb}$ indicates the exoplanet's orbital velocity around its host star, $\phi$ represents the orbital phase of the exoplanet, $i$ refers to the inclination of the system relative to the observer's line of sight, and $v_{sys}$ corresponds to the systemic velocity of the host star.

We shift the cross-correlation values according to the equations for a presumed semi-major amplitude ($
K_p$, which is $v_{\text{orb}} \times \sin{i}$ in Eq. \ref{eq:orbital_equation}) by applying the known orbital phase of the planet. To find the optimal alignment, we sample a range of $K_p$ values. Each one is then collapsed along the time axis by computing a vertical average. When the shift leads to vertical alignment of the cross-correlation function, a peak becomes evident. However, if the alignment is not accurate, the cross-correlation functions will not constructively interfere with each other. As a result, we expect to see a peak at the correct $K_p$ value.

We construct a grid, or a \kpvsys\ map~\citep{Brogi_2012}, from all the collapsed cross-correlation functions corresponding to the various shifts. This map is identified by a peak at the correct $K_p$ and $v_{sys}$ values. The row corresponding to this peak provides an optimised one-dimensional CCF, where all exposures are combined to yield the most robust signal. For this study, we used a grid of semi-major amplitudes spanning from 0 to 400 \kms, incremented in steps of 1 \kms.

Finally, we extract the one-dimensional CCF by selecting the $K_p$ that best aligns the CCF functions. In order to determine the amplitude, width, and centre locations of the signal, we employ a least-squares fit to the one-dimensional CCF under the assumption of a Gaussian profile. Uncertainties in the amplitude, width, and centre locations are derived via a bootstrap method, which involves sampling the noise floor of the one-dimensional CCF\@.

This process begins by removing a region in the one-dimensional CCF where the peak is located, specifically for CCF values ranging from $-50$ \kms to $10$ \kms. We then compute the standard deviation of the remaining values to estimate the standard deviation of the noise. Assuming a normally distributed noise profile based on this computed standard deviation, we create a Gaussian noise model. We then bootstrap all points in the CCF, randomly altering each point using this noise model, applying the same least-squares Gaussian fit, and recording the measured parameters each time. After repeating this process for 5,000 iterations, we calculate the standard deviation of the fitted values, treating these as their uncertainties. The statistical significance of the signal is then evaluated as the ratio of the measured amplitude to its corresponding uncertainty.


\begin{figure*}[ht]
\centering
\includegraphics[width=\textwidth]{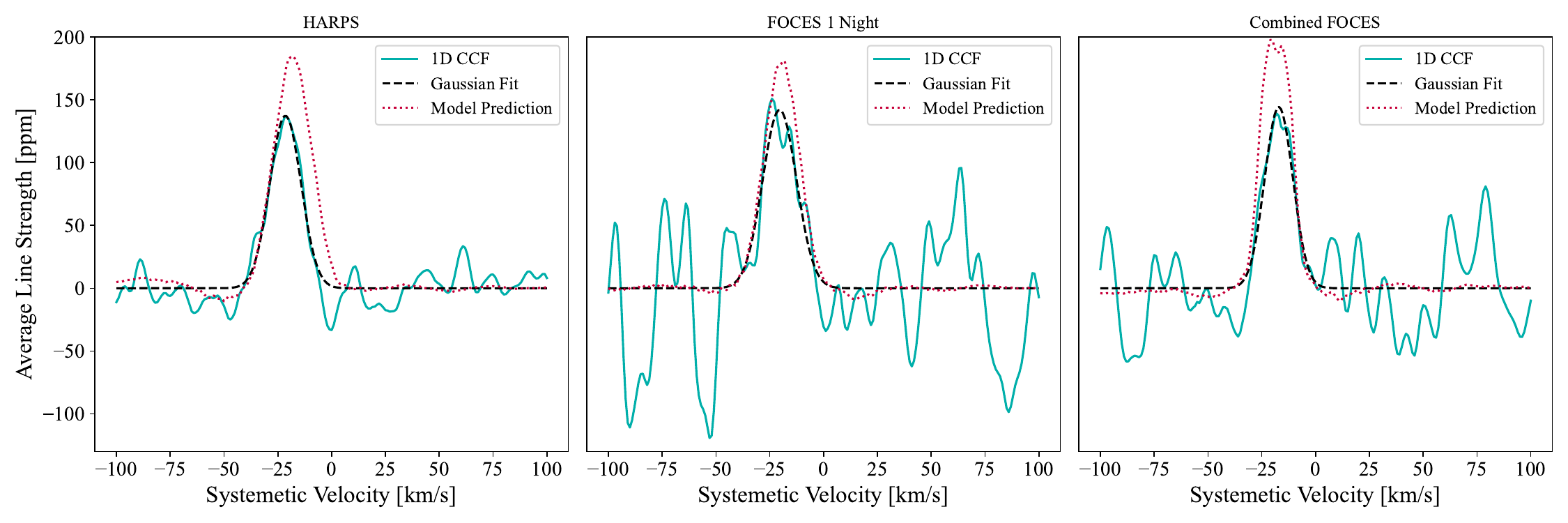}
\caption{Comparison of 1D CCFs for the Fe\,I signal between different datasets extracted from the peak of each \kpvsys plot. Each panel represents the data from a different source: HARPS-N, FOCES Night 1, and FOCES nights 1, 2, and 3 combined, respectively. In each plot, the teal line represents the one-dimensional cross-correlation function, the black dashed line is the Gaussian fit to the peak in the plot, and the red dotted line is the model prediction, as discussed in Sect. \ref{subsec:model_injection}. There is a small discrepancy in the model peak height for each result; this is likely driven by the larger wavelength range of FOCES, which changes the predicted cross-correlation result. The model-predicted CCF has values of slightly below zero near the peak of the function. This observation is attributed to cross-correlation aliases, a phenomenon explained in \cite{Borsato_2023}.}
\label{fig:oned_modelcomparison}
\end{figure*}

\begin{table*}
\caption{Comparative statistics of CCF from the HARPS-N and FOCES observations for the Fe\,I detection}
\centering
\begin{tabular}{lccc}
\hline \hline
Parameter & HARPS-N & FOCES & Combined FOCES \\
\hline
Amplitude [ppm] & $132.7 \pm 6.9$ & $142.1 \pm 24.4$ & $123.1 \pm 17.2$ \\
Centre \big[\kms\big] & $-21.3 \pm 0.5$ & $-20.4 \pm 1.6$ & $-16.4 \pm 1.7$ \\
Width \big[\kms\big] & $7.0 \pm 0.4$ & $7.9 \pm 1.5$ & $8.2 \pm 1.2$ \\
Detection significance & $19.2$ & $5.8$ & $7.2$ \\
Model discrepancy & $0.3 \pm 0.04$ & $0.1 \pm 0.1$ & $0.3 \pm 0.1$ \\
Discrepancy significance & $7.5$ & $1.0$ & $3.0$ \\
\hline
\end{tabular}
\tablefoot{This tables shows the summary statistics from the bootstrap fits of the one-dimensional cross-correlation function for Fe\,I. The column labelled `FOCES' showcases the fit for the most successful observation night, which was on 18 July 2022. Furthermore, the `Combined FOCES' column presents a fit applied to results generated by following the methodology described in Sect. \ref{subsec:combining_nights}. Detection significance is the ratio of the amplitude fits to their measured standard deviation. Model discrepancy is the measured difference between the predicted signal of the model and the actual signal. It is measured as one minus the ratio of the fitted amplitude of the cross-correlation function against the height of the model's peak.}
\label{ComparisonTable}
\end{table*}

\subsection{Combining the FOCES nights}
\label{subsec:combining_nights}
To amalgamate the three FOCES observations into a single \kpvsys\ map, we employed an approach adapted by~\cite{Borsato_2023}, which involves computing a weighted average of the \kpvsys\ maps. Given that only one spectrograph is under consideration, we do not need to factor in the instrument's wavelength range, which is part of the process used in this latter publication. We therefore proceed by constructing appropriate weights to combine the nights.

Firstly, we mask out the signal from the planet in the \kpvsys\ map. Our mask was devised to screen all data values ranging from $-50$ \kms to $10$ \kms for the systemic velocity, and $100$ \kms to $300$ \kms for the orbital velocity, which is the same region applied in \cite{Borsato_2023}. The selected window intentionally spans a broad velocity range to capture all known atmospheric detections of exoplanets, while allowing for possible offsets arising from atmospheric and dynamical effects \citep{Prinoth_2022}. Additionally, changes in the region boundaries do not have a large effect on results provided the signal is contained within it. Upon applying the mask, we calculate the standard deviation of the remaining pixels, providing a measure of any residual noise and systematic uncertainties affecting the overall noise floor.

Next, we invert the mask, thereby isolating the region where we anticipate the signal. Any signal within this location means that the region will yield a higher average value compared to the background. By taking the ratio of this mean to the calculated standard deviation, we generate a pseudo-S/N, which offers a general assessment of the quality of the night in terms of S\textbackslash N and can then be converted into the desired weight. Lastly, we normalise the pseudo-S/N values for all observations so that they add up to one, using these resultant values as the weights.

\subsection{Cross-correlation templates}

In this study, we used the standard templates provided by the Mantis network~\citep{Kitzmann_2021}, which comprises more than 750 high-resolution templates for various species across a grid of different atmospheric temperatures. These templates are publicly available\footnote{\url{https://cdsarc.cds.unistra.fr/viz-bin/cat/J/A+A/669/A113}} and are computed using a forward modelling approach, taking into account the transmission spectrum of the target planet and the absorption lines of the species under scrutiny. The computation process includes opacity functions, continuum opacity contributions, and assumptions about metal abundance and temperature profiles. These templates provide a standardised and efficient alternative to creating new templates for each individual study. Mantis templates have been successfully applied in the detection of a range of species using the ESPRESSO spectrograph on the VLT~\citep{Silva_2022}. Our search only focused on the species reported in~\cite{Hoeijmakers_2019} ---which have an assumed equilibrium temperature of 4000 K--- to simplify the comparison between telescopes.

\subsection{Model injection and comparing instruments}
\label{subsec:model_injection}

In order to make a comparative analysis of the data from FOCES and HARPS-N, our initial step was to create a forward model of KELT-9\,b's transmission spectrum, taking inspiration from similar studies~\citep[e.g.][]{Hoeijmakers2019AB}. We used the same model as described by \citet{Kitzmann_2023} to compute the transmission spectrum but changed the planet's radius and surface gravity to that of KELT-9\,b. The temperature profile of the atmosphere was assumed to be isothermal, with a value of 4000 K\@.
We used all opacity species listed in Table 2 of \citet{Kitzmann_2023}. The corresponding line absorption coefficients were calculated with the HELIOS-K\footnote{\url{https://github.com/exoclime/HELIOS-K}} opacity calculator~\citep{Grimm_2015, Grimm2021ApJS..253...30G}. The chemical composition of the atmosphere was obtained using the equilibrium chemistry code \textsc{FastChem 2}\footnote{\url{https://github.com/exoclime/FastChem}}~\citep{Stock_2018,Stock_2022}, assuming solar element abundances. The resultant transmission spectrum, which is representative of the observed atmospheric characteristics, is shown in Fig. \ref{fig:model_for_injection}.

Following the methodology outlined in~\cite{Hoeijmakers_2015} (which searched for TiO in HD 209458 b, and used model injection to explain its non-detection), we take the predicted transmission spectrum, inject it into the pipeline-reduced spectra obtained from the observations of both instruments, and rotationally broaden the modelled lines to account for the planet's rotation. Subsequently, we perform a Gaussian convolution to match the spectral resolution of each spectrograph. The wavelength axis is
interpolated in order to align with the spectrograph grid. Finally, we perform cross-correlation on the spectra a second time. This approach facilitates a comparison between the expected cross-correlation signal and the actual signal generated by the spectrograph. By subtracting the cross-correlation results for the case with an injected signal from the cross-correlation profile without an injected signal we obtain a line depth and shape that are predicted by the model. These are shown in Fig. \ref{fig:oned_modelcomparison}. We then use this predicted line to calculate the model discrepancy, as detailed in Table~\ref{tab:ObservationalSummary}.

\section{The detection feasibility of a 2m telescope}
\label{sec:feasability_of_2m_telescope}
While our observations focus on the results from observations made at the Wendelstein Observatory, we aim to expand our analysis to consider 2m class telescopes in general. Our goal in this section is to develop a transit emulator\footnote[2]{\url{https://github.com/nborsato/transit_emulator}} that generates mock cross-correlation results of transit spectra. To do this, we use measurements of the S/N of Wendelstein as a baseline to extrapolate the capability of 2m class telescopes in observing exoplanet atmospheres. For the best Wendelstein night, the $\textrm{S/N}\sim35$ for the host-star KELT-9, which has a visual magnitude of 7.6. Taking this as an ideal S/N, we can create an approximate analytical equation that estimates the S/N for a specific stellar magnitude:

\begin{equation}
    S/N \approx 35\sqrt[]{10^{(m-7.6)}}
    \label{eq:snr_eq}
,\end{equation}

where $m$ is the magnitude of the star in question. We require a model in order to evaluate the detectability of the atmosphere of a UHJ. In this case, we selected spectral templates for Fe\,I from the Mantis template library of~\cite{Kitzmann_2021} and used these as models of the atmosphere of UHJs in general. We chose these templates over our own generated model to simplify the approach and increase reproducibility. To simulate the transit of a UHJ, we first reduced the templates down to a resolution of ~70,000 by applying Gaussian convolution. For this general case, we used the KELT-9\,b transit duration of 3.91 h~\citep{Gaudi_2017}, which we divided into steps of 330 s to simulate an exposure time of 300 s plus 30 seconds of overhead readout.

With each step, we calculate the phase angle of the planet:
\begin{equation}
    \phi = \frac{2\pi(t)}{p}
,\end{equation}

where $t$ is the time step in question, which has a value range of $t \in \left[-\frac{3.91}{2}, \frac{3.91}{2}\right] \, h$, and $p$ is the orbital period of the planet. With orbital velocity, phase angle, and systemic velocity obtained from~\cite{Gaudi_2017}, we apply

\begin{equation}
    v_{r} = v_{orb}\sin{\phi} + v_{sys}
\end{equation}

to obtain the radial velocity of the planet at each exposure time, and we then create a time series to emulate the exoplanet transit by Doppler shifting the template for each calculated radial velocity value. This results in 42 exposures for the entire transit. We also calculate the differences in the radial velocity steps, and further broaden the spectral lines of exposure to account for the line broadening caused by the Doppler shift in radial velocity space that the planet undergoes through each exposure.

Finally, we add Gaussian noise with a mean of zero to each spectrum using the S/N estimate provided by Eq. \ref{eq:snr_eq}, converting it to a standard deviation. This process creates a set of mock transit exposures for which we can perform cross-correlation. We then follow the procedure outlined in Sect. \ref{sec:ccfapproach} to constrain detection significance.

\begin{figure}[h]
    \centering
    \includegraphics[width=\columnwidth]{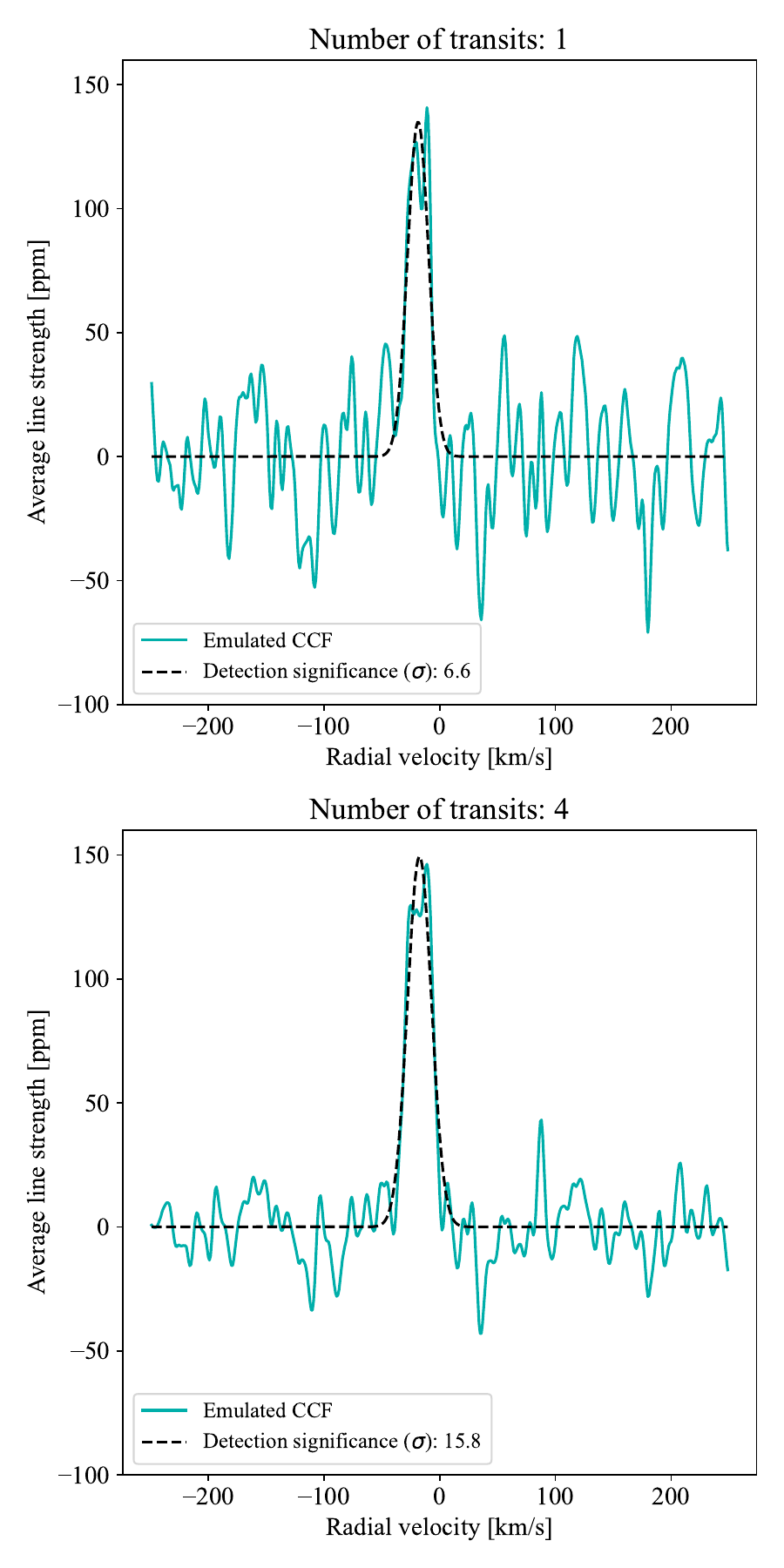}
    \caption{Signal enhancement through transit stacking as demonstrated by the emulator developed in this study. Both panels depict the one-dimensional cross-correlation function (teal) alongside the Gaussian fit to the line profile (black dashed lines) for a single night, with the significance of the fit specified in the legend. The top panel illustrates the results for one night, whereas the bottom panel shows the amplified signal clarity achieved when four transits are stacked together, underscoring a marked reduction in the noise profile. This comparison demonstrates the benefits of transit stacking for improved signal detection.}
    \label{fig:emulator_results}
\end{figure}

With our mock-transit spectrum generator, we first emulated a transit of KELT-9\,b using the setup presented above to compare how well our process matches the observations. Further, we took an average of four emulated nights to determine the extent to which we should expect the statistical significance to increase when combining nights, and to determine whether or not combining four transits will produce the same results as a 4m class telescope. Finally, with our understanding of the performance of this emulator, we emulated observations using Fe\,I templates at equilibrium temperatures of 2500\,K, 2000\,K, and 4000\,K for stars of different magnitude, ranging from 6 to 11.5 in steps of 0.5 mag. We used three templates in order to to consider the effect that equilibrium temperature has on detection significance and to gain a better idea of the potential observability of the planets put forward in Table~\ref{tab:potential_targets}. We emulated 30 transits for each magnitude and took combined averages over an increasing number of nights, up to 30, observing the increase in statistical significance with the addition of each new transit.

\begin{figure*}[h]
    \centering
    \includegraphics[width=\textwidth]{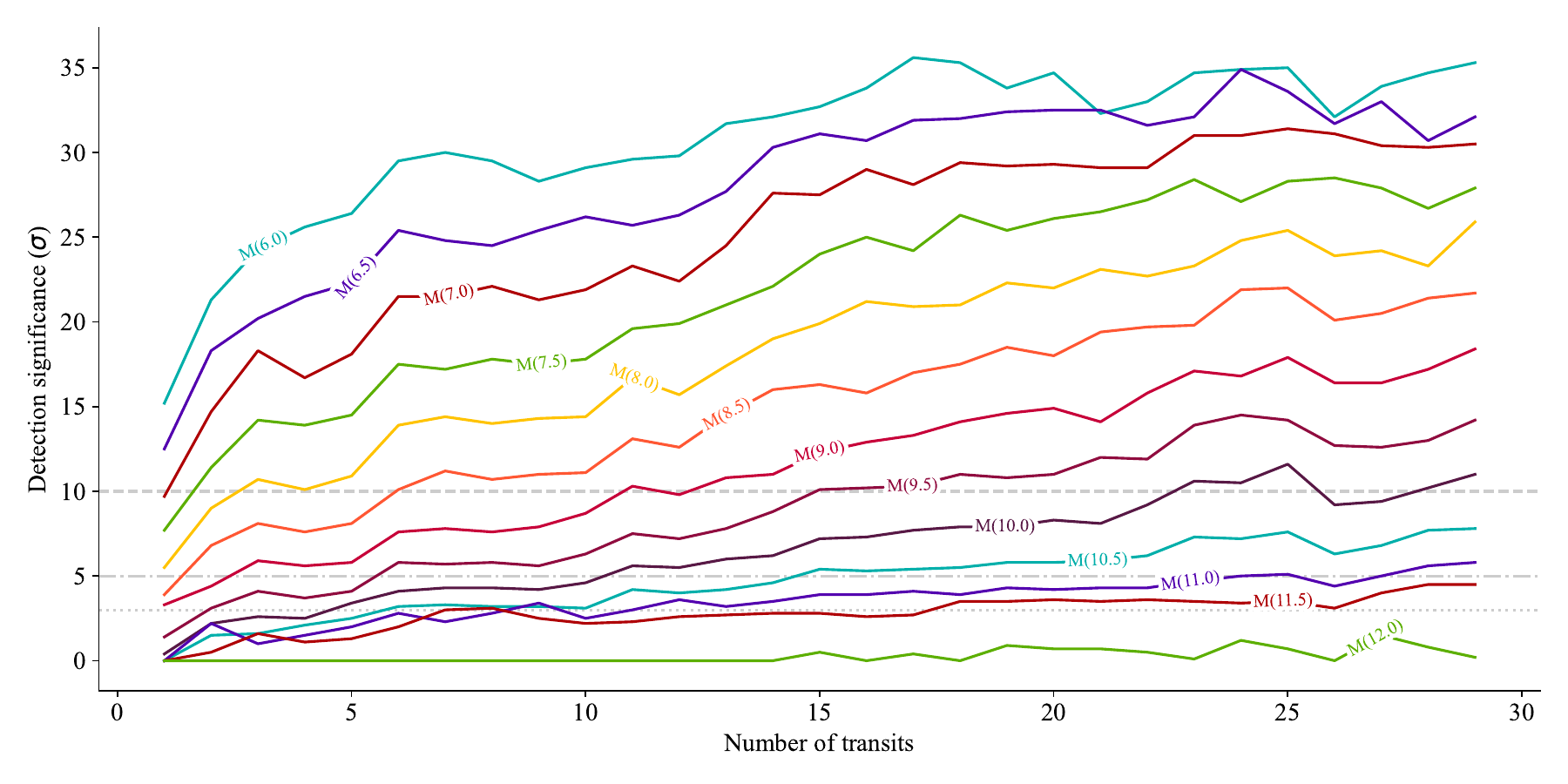}
    \caption{Detection significance trends when stacking transit observations based on the S/N of Wendelstein. Each line represents the number of transits required to reach the 3, 5, and 10$\sigma$ significance thresholds marked as the grey dashed horizontal lines. These predictions are based on the 2500\,K template, which has shallower lines, and therefore these predictions are conservative and could be more robust on a per-transit basis if the equilibrium temperatures of the planets were higher.}
    \label{fig:transit_significances_trend_2500}
\end{figure*}
\normalfont

\section{Results}
We first present a comparative visual representation of the Fe\,I detection from a single night using HARPS-N and the first night with FOCES. We illustrate these results in Fig. \ref{fig:cross_correlation_results_one_night}, where in the upper portion of the plots we show the cross-correlation function before conversion to a \kpvsys\ map. In the centre of these plots, we present the \kpvsys\ map, which shows the signal of Fe\,I for both spectrographs; though the HARPS-N signal is much stronger. Finally, in the lower portion of these figures, we present the extracted one-dimensional CCF, which shows the idealised signal produced from the cross-correlation process. We are able to detect atomic iron using both spectrographs, as reflected in the \kpvsys plots and the one-dimensional CCFs. However, a prominent difference emerges in the trace of the planet within the CCF: HARPS-N data clearly reveal the trace, whereas it is only partially discernible in the FOCES output. A closer look at the one-dimensional function illustrates the dominance of noise in the FOCES data for that night. Nevertheless, the emergence of the signal confirms the feasibility of resolving the atmosphere of KELT-9\,b with a single transit, even when baseline out-of-transit observations are minimal.

In Fig. \ref{fig:oned_modelcomparison}, we compare the performance of the FOCES to that of the HARPS-N spectrograph for a single night of observation, and when all three nights are combined. We do this by plotting the one-dimensional CCFs with their corresponding Gaussian fits, and the signal predicted by the model as described in Sect. \ref{subsec:model_injection}. The values and statistics for each fit are provided for comparison in Table~\ref{ComparisonTable}, including how the model compares to the actual data. At first glance, it is clear that HARPS-N provides superior signal quality even when combining the transits. However, the line depths are comparable across all three nights. When comparing to the model injection with the observations depicted in Fig. \ref{fig:oned_modelcomparison}, the discrepancy, computed as the difference between the signal and the model, is approximately equivalent for both the HARPS-N observation and the combined FOCES observations, although the uncertainties associated with the FOCES discrepancy are nearly twice as large as those produced by HARPS-N. This implies that, when juxtaposed with our model, the signals derived from both spectrographs diverge from the model to a similar extent.

\begin{figure*}[h]
    \centering
    \includegraphics[width=\textwidth]{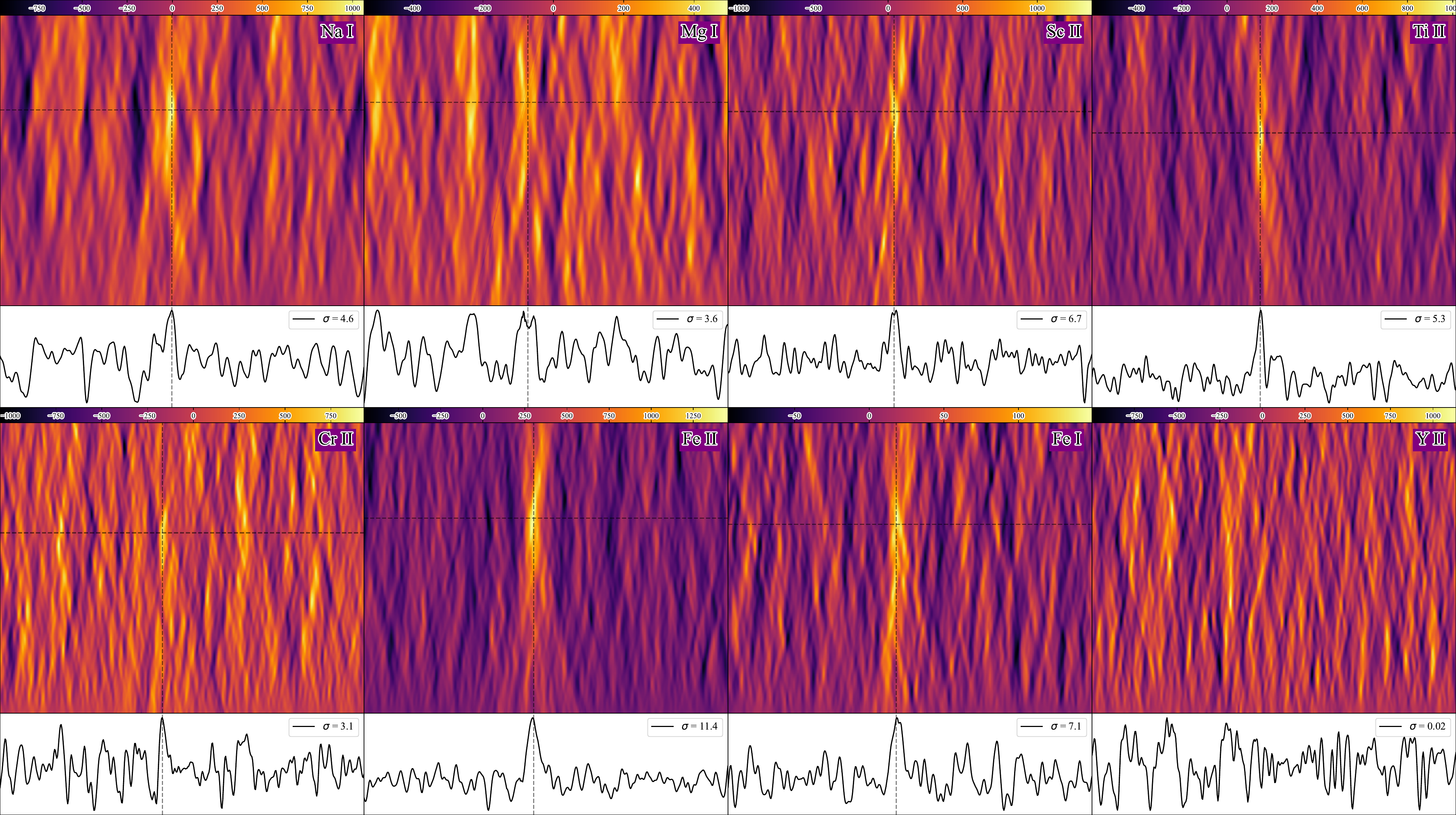}
    \caption{Cross-correlation results for the combined FOCES observations across the eight species investigated. Ordered left to right, then top to bottom are: Na\,I, Mg\,I, Sc\,II, Ti\,II, Cr\,II, Fe\,II, Fe\,I, and Y\,II. The top panels of each detection contain the \kpvsys\ maps. The cross-hairs locate the points where the signal peak has been extracted, and the bottom plots show the one-dimensional cross-correlation slice after extracting the rows containing the peak. The detection significance ($\sigma$) is indicated in the legend. We have managed to successfully recover seven of the eight detections reported in \cite{Hoeijmakers_2019}. Only Y\,II has not been recovered, but it is known to be a weak signal and difficult for the HARPS-N spectrograph to resolve. While the signals are present in most plots they exhibit a strong degree of noise, indicative of using a 2m class telescope.}
    \label{fig:all_detections}
\end{figure*}

The \kpvsys maps and one-dimensional CCFs for the combined FOCES nights, across all eight detections, are displayed in Fig. \ref{fig:all_detections}. The amplitude, width, centre locations, and detection significance are detailed in Table~\ref{Detections_Table}. Using our method, we were able to successfully replicate seven out of the eight detections previously mentioned in~\cite{Hoeijmakers2019AB}. The only exception was Y\,II, which we did not find in our results. This is consistent with the difficulties encountered in reanalysing the HARPS-N data to confirm the presence of Y\,II, as noted by~\citet{Borsato_2023}. If Y\,II is indeed present, its relatively weak signal is likely due to its low abundance in the planet's atmosphere. More definitive conclusions about the presence of Y\,II would require further observations, either through combined efforts or with the use of a larger telescope.

Among the detected signals, Fe\,II appeared to be the strongest. However, we also noted substantial detections of Fe\,I and Ti\,II, along with weaker detections of Na\,I, Mg\,I, Na\,II, Cr\,II, and Sc\,II. The central peaks of these signals were measured using a least-squares Gaussian fit, which estimates the best fit for the centre, width, and amplitude of the cross-correlation peak. The derived parameters aligned with the findings of~\cite{Hoeijmakers_2019}, with the uncertainties of our measurements overlapping with those reported by these authors. However, the tentative Mg\,I detection merits particular attention; it exhibits considerable smearing and is significantly delocalised from its actual signal, implying that there may be additional issues with this detection, and that further exploration is warranted.

We now move on to review the results of our transit emulator. Referring to Fig. \ref{fig:emulator_results}, we see that our emulation yields a significance value of
6.6$\sigma$ for a single emulated transit, which compares favourably with the FOCES value of 5.8$\sigma$ in Table \ref{ComparisonTable}, indicating a slightly higher significance. This discrepancy likely stems from the fact that the S/N drops towards the end of the observation, while in the emulated case, it remains consistent. When averaging over four nights, the significance increases to 15.8$\sigma$, which aligns well with a single HARPS-N night; although the significance is slightly lower, it represents a much larger improvement compared to the stacked observations produced by the FOCES spectrograph. Our results from running the emulator with the 2500\,K template are depicted in Fig \ref{fig:transit_significances_trend_2500}. We note that single-transit detections are only possible at a 5$\sigma$ significance level for magnitudes of 7.5 or lower, while dimmer targets necessitate multiple observations. However, even planets orbiting around stars of 10.5 mag attain a constrained 5$\sigma$ detection if approximately 20 transits are captured, suggesting that all the UHJs listed in Table~\ref{tab:potential_targets} could feasibly be observed with a 2m class telescope. A final point to note is that detection significance appears to plateau, particularly for the brighter targets; for instance, the magnitude 6.0 case stabilises around a detection significance of 20$\sigma$ after about 7 transits. We discuss this point further in the following section.

\begin{table*}
    \centering
    \caption{Summary of the orbital and physical characteristics that could be observable with FOCES}
    \tiny 
    \begin{tabularx}{\textwidth}{XXXXXXXXX}
    \hline \hline
    Name & RA (deg) & Dec (deg) & $V_{\text{mag}}$ & $T_{\text{eq}}$ (K) & Period (days) & Semi-major axis (AU) & Mass ($M_{\text{Jup}}$) & Inclination (deg) \\
    \hline
    KELT-20\,b & 294.663 & 31.219 & 3.0 & 2230.0 & 3.5 & 0.054 & 3.5 & 86.2 \\
    WASP-189\,b & 225.687 & -3.031 & 6.6 & 3353.0 & 2.7 & 0.051 & 2.0 & 84.03 \\
    KELT-9\,b & 307.86 & 39.939 & 7.6 & 3921.0 & 1.5 & 0.034 & 2.9 & 87.2 \\
    TOI-1431\,b & 316.204 & 55.588 & 8.0 & 2370.0 & 2.7 & 0.047 & 3.1 & 80.4 \\
    KELT-7\,b & 78.296 & 33.318 & 8.5 & 2048.0 & 2.7 & 0.045 & 1.2 & 83.9 \\
    HD\,202772\,A\,b & 319.7 & -26.616 & 8.3 & 2132.0 & 3.3 & 0.052 & 1.0 & 84.2 \\
    KELT-17\,b & 125.617 & 13.735 & 9.3 & 2087.0 & 3.1 & 0.049 & 1.3 & 84.87 \\
    HAT-P-70\,b & 74.55 & 9.998 & 9.5 & 2562.0 & 2.7 & 0.047 & 6.8 & 96.5 \\
    HD 2685\,b & 7.329 & -76.304 & 9.6 & 2061.0 & 4.1 & 0.057 & 1.2 & 89.4 \\
    KOI-13 b\,& 286.971 & 46.868 & 10.0 & 2550.0 & 1.8 & 0.036 & 9.3 & 86.77 \\
    TOI-2109\,b & 253.188 & 16.58 & 10.0 & 3631.0 & 0.7 & 0.018 & 5.0 & 70.74 \\
    \hline
    \end{tabularx}
    \label{tab:potential_targets}
    \tablefoot{The table includes right ascension (RA), declination (Dec), visual magnitude ($V_{\text{mag}}$), equilibrium temperature ($T_{\text{eq}}$), orbital period, semi-major axis, mass (in terms of Jupiter mass $M_{\text{Jup}}$), and orbital inclination.}
\end{table*}

\section{Discussion}
In ideal observing conditions, the HARPS-N spectrograph on the TNG can resolve the atmosphere of KELT-9\,b with impressively high fidelity with a single transit observation. However, our results ---shown in Fig \ref{fig:cross_correlation_results_one_night}--- demonstrate that FOCES, on a telescope half the size of the TNG, can also resolve the atmosphere of this source even when observation conditions are poor. We therefore propose that by using multiple transits, results comparable to those achieved with HARPS-N can also be realised using moderately sized telescopes provided that the stellar signal can be resolved and that a sufficient number of transits can be obtained.

The HARPS-N spectrograph outperforms the FOCES spectrograph by a factor of two to three considering all three combined nights of FOCES observations. This disparity is larger than anticipated based solely on telescope size; we would have expected four nights to be sufficient to yield detection statistics comparable to those of HARPS-N from one night. Therefore, there are other factors degrading the quality of these observations.

The prevailing weather conditions, particularly cloud cover, were likely the most substantial factors influencing the results. The cloud cover during all FOCES observations likely had a substantial impact on the detection statistics used for the comparison. The presence of clouds at some point during each of the three observations hindered the process of making a robust comparison between the two telescopes and their respective spectrographs.

An alternative explanation for the deteriorated data quality could be the air mass during the observations. However, when comparing all the observations in Fig. \ref{fig:airmassplots}, we observe that all were conducted in conditions of low air mass, suggesting this variable likely did not contribute significantly to the discrepancy in the results. A portion of all the three transits observed by FOCES took place during astronomical twilight, potentially introducing contamination from diffuse background light. However, as this contamination would not consistently affect all exposures, its overall impact is likely to be minimal.

An alternative factor that may have played a smaller role is the limited baseline measurements of the host star. The single night of HARPS-N has baseline observations that almost match the total baseline duration of all three FOCES observations. Despite the absence of additional planetary flux, this increase in the measurement of the stellar flux could permit a more precise constraint on the stellar component subtracted during the cross-correlation process. To draw more definitive conclusions, additional controls are necessary, including consistent baseline timings, similar observing windows, and most importantly comparable weather conditions.

Despite these issues, our findings show that FOCES at the Wendelstein Observatory is capable of detecting signals from a multitude of chemical species present in the atmosphere of KELT-9\,b. As detailed in Table~\ref{Detections_Table} and Fig. \ref{fig:all_detections}, we successfully retrieved spectral signals of seven atomic and ionised species, each with a detection significance exceeding 3$\sigma$. Moreover, the enhancements in detection significance, indicated in Table~\ref{ComparisonTable} and Fig. \ref{fig:oned_modelcomparison}, suggest that multiple transit observations can lead to results that are of comparable quality to those from HARPS-N. We note that HARPS-N has successfully conducted day-side observations of KELT-9\,b~\citep{Pino_2020}. Based on our findings, we believe that it is reasonable to hypothesise that analogous observations could be carried out using FOCES. These results suggest that smaller telescopes could play a pivotal role in broadening the scientific scope of exoplanet atmospheric studies.

The distinctive advantage that small telescopes offer lies in the abundance of their observational time. These telescopes are capable of revisiting targets repeatedly, which not only places them in a competitive position with larger classes of telescopes, but also allows them to track the temporal evolution of these planetary atmospheres. Larger telescopes, due to intense competition for observation time and the drive for unique discoveries, tend to impose limitations on the scope of potential scientific questions that may require substantial amounts of observing time. Therefore, small telescopes could play a crucial role in the systematic monitoring of such planetary atmospheres, especially considering their known dynamic evolution. For example, recent observations suggested that KELT-9\,b is gradually losing its atmosphere~\citep{Wyttenbach_2020}. Therefore, the persistent and repetitive observational capabilities of small telescopes could provide invaluable insights into the evolution of the atmospheres of exoplanets.

Leveraging the successful detections achieved with FOCES, we generalised the observational capabilities of 2m telescopes through the creation of an emulator. The emulator reproduces a similar statistical significance to FOCES, and also creates a statistical significance comparable to HARPS-N when four emulated nights are combined, as shown in Fig. \ref{fig:emulator_results}. This outcome suggests that resolving the atmospheres of exoplanets only requires stacking transits and can yield results comparable to those obtained with larger classes of telescopes. With this success, we generalised the average S/N of FOCES to predict how many transits would be required to resolve the atmospheres of other known UHJs with a 2m telescope. Using the exoplanet archive, we compiled a list of potential UHJ targets that could feasibly be observed with a 2m telescope in under 20 transits, as shown in Table~\ref{tab:potential_targets}. Such a high number of observations would be a formidable undertaking for a large telescope, but is indeed plausible given the ample observation time available with 2m telescopes. We chose targets that exhibit a sufficiently high atmospheric equilibrium temperature to allow molecular dissociation, thereby enabling us to observe atomic and ionised line transitions at optical wavelengths. Should an observation programme be launched with 2m telescopes, most targets would need to be observed over several transits, except for planets with magnitudes of less than 7.5 (see Figs. \ref{fig:transit_significances_trend_2500}, \ref{fig:transit_significances_trend_2000}, and \ref{fig:transit_significances_trend_4000}). Admittedly, unforeseeable factors, such as weather, may lead to the requirement for additional observations in order to offset lost nights. Nevertheless, with perseverance, 2m telescopes, like that at the Wendelstein Observatory, have significant potential for the observation of an extensive range of UHJ atmospheres.

In addition to providing arguments for the potential of 2m telescopes, our emulator seems to suggest there is a limit to the increase in statistical significance when stacking transit observations. In Fig. \ref{fig:transit_significances_trend_2500}, for stars with visual magnitudes of 6.5, significance increases with the square root of the number of transits, yet appears to reach a limit after multiple transits have been stacked; this trend can also be seen in Fig. \ref{fig:transit_significances_trend_4000}, and to a lesser extent in \ref{fig:transit_significances_trend_2000}. However, the limit lies at different levels of significance: the 4000\,K template reaches a significance of 50, while the 2500\,K template only reaches a significance of 25. To better illustrate this phenomenon, we present Fig. \ref{fig:divergance_from_poisson}, which compares the increasing trends in detection significance for emulated stars of visual magnitudes 6.0 and 11.5. This plot distinctly reveals a divergence from the expected Poisson trend in the case of the magnitude 6.0 star. The explanation we propose is that with a sufficient number of transits, the alias floor of the CCF \citep{Borsato_2023} becomes resolved, becoming the primary contributor to the standard deviation calculation used to model the noise profile described in Sect. \ref{sec:feasability_of_2m_telescope}, causing the noise level to stay locked at this detection significance. In this case specifically, the aliases arise only from the self-correlation of lines in the Fe\,I template. In an actual case, other additional aliases can come from different species in the planetary spectrum and could have a stronger effect. The reason for this variation in level is linked to the differences in the line profiles of each template, which vary in the amount of aliasing lines present in the CCF. Figure \ref{fig:equib_templates_comparison} illustrates this, showing that more lines are present in the 4000\,K template than in the 2500\,K and 2000\,K templates, respectively. This result suggests that the HARPS-N observation of KELT-9 b is likely already resolving the alias floor, with a single observation being sufficient to constrain the noise profile. Therefore, in high-S/N regimes, stacking transits yields diminishing returns with the cross-correlation technique, unless aliasing is accounted for.

While FOCES provides a demonstration of the potential of a high-performance spectrograph on a 2m telescope, its scientific reach is primarily limited to UHJs, because its wavelength range is limited to optical wavelengths. Recent studies, particularly those involving the GIANO-B spectrograph on the TNG~ \citep{Brogi_2018, Guilluy_2019}, have convincingly demonstrated the scientific utility of high-resolution near-infrared observations for exoplanets with lower equilibrium temperatures, such as hot Jupiters. This suggests that similar science cases—ranging from helium evaporation~\citep{Oklo_2018,Nortmann_2018} to the detection of various molecular species, such as CO~\citep{Yan_2022}--- could potentially be replicated on smaller telescopes, provided they are outfitted with comparable near-infrared capabilities. The capability for long-term baseline measurements facilitated by small telescopes could offer in-depth insights into this evaporation process, potentially shedding light on competing theories of planetary evaporation, such as photo-evaporation~\citep{Hu_2015,Owen_2017} versus core-powered mass loss~\citep{Ginzburg_2018,Gupta_2019,Gupta_2020}. Given the comparable performance of FOCES with respect to that of HARPS-N, it is plausible that a similar near-infrared spectrograph like GIANO-B could achieve similar results.

Large telescopes often excel in scientific investigations that demand precision and reduced noise levels; for example when resolving individual lines~\citep[e.g.][]{Seidel_2022,Asnodkar_2022}, analysing atmospheric dynamics~\citep{Ehrenreich_2020,Prinoth_2022}, or hunting for fainter signals from heavier elements~\citep{Jiang_2023}. However, a common challenge with large telescopes is their oversubscription. It is assumed that oversubscription leads to prioritisation of the best-qualified projects~\citep{DeCastro_2003}; which while generally accurate, reduces the scope for creative experimentation and diverse observational strategies. It is worth noting that the first exoplanet discovery, 51 Pegasi b, was made using a 2m class telescope~ \citep{Mayor_Queloz_1995}, despite the availability of larger telescopes at the time. Our study demonstrates that when it comes to studying exoplanet atmospheres, smaller telescopes can deliver comparable performance to their larger counterparts, enabling further experimentation and possibly leading to new discoveries.

\section{Conclusion}
In this study, we successfully resolved the atmosphere of KELT-9\,b in a single transit, and detected seven of the eight detections presented in~\cite{Hoeijmakers2019AB}. Additionally, we compared the performances of the high-resolution spectrograph FOCES at the Wendelstein Observatory to the HARPS-N spectrograph on the TNG and found promising evidence that, given more observations, FOCES can produce results of a comparable level of quality to those obtained with HARPS-N. UHJs can be viewed as testing grounds for observational concepts related to atmospheres. With our results it can be argued that smaller telescopes carry the potential to explore new ideas and reframe the scientific questions that we should pursue with larger telescopes. Our results demonstrate that efforts to maintain smaller, high-performance telescopes are worthwhile, that 2m telescopes are likely to continue to deliver pioneering science, and suggest that we should even expand the instrumentation they are equipped
with in order to study UHJ atmospheres in greater detail.

\begin{acknowledgements}
       N.W.B acknowledge partial financial support from The Fund of the Walter Gyllenberg Foundation. BE and KM are supported by the Excellence Cluster ORIGINS which is funded by the Deutsche Forschungsgemeinschaft (DFG, German Research Foundation) under Germany’s Excellence Strategy - EXC-2094-390783311.
\end{acknowledgements}

\bibliographystyle{aa} 
\bibliography{bib}

\appendix
\section{Detections}
We present the measured parameters of each cross-correlation peak for each signal recovered when combining the three FOCES observations. Table~\ref{Detections_Table} outlines the measured amplitude, centre, width, and detection significance of each cross-correlation peak. All detection significances, except for Y\,II, are detected above a 3$\sigma$ threshold.

\begin{table*}[h]
\centering
\caption{Summary of the detection parameters and significance for the each of the eight species in the transmission spectrum of KELT-9\,b. }
\begin{tabular}[width=\textwidth]{lcccc}
\hline\hline
Element & Amplitude [ppm] & Centre \big[\kms\big] & Width \big[\kms\big] & Detection Significance \\
\hline
Na\,I & $1118 \pm 242$ & $-17.1 \pm 2.49$ & $5.83 \pm 2.31$ & 4.6 \\
Mg\,I & $403 \pm 112$ & $-26.8 \pm 2.68$ & $7.30 \pm 1.60$ & 3.6 \\
Sc\,II & $1281 \pm 191$ & $-21.7 \pm 1.15$ & $5.29 \pm 1.06$ & 6.7 \\
Ti\,II & $976 \pm 183$ & $-19.1 \pm 1.64$ & $5.06 \pm 2.75$ & 5.3 \\
Cr\,II & $986 \pm 322$ & $-25.9 \pm 1.57$ & $3.02 \pm 1.56$ & 3.1 \\
Fe\,II & $1387 \pm 121$ & $-17.4 \pm 0.81$ & $7.65 \pm 0.87$ & 11.4 \\
Fe\,I & $147 \pm 21$ & $-16.9 \pm 1.25$ & $6.62 \pm 1.09$ & 7.1 \\
Y\,II & $20 \pm 1080$ & $-20.2 \pm 5.48$ & $12.4 \pm 232$ & 0.02 \\
\hline
\end{tabular}
\tablefoot{The table lists the amplitude, centre, width, and detection significance for each detected element}
\label{Detections_Table}
\end{table*}

\section{Air-mass plots}
Here, we present the air-mass plots for each night of observation, highlighting where the transit occurs.

\begin{figure*}
    \centering
    \subfloat{
        \includegraphics[width=0.49\textwidth]{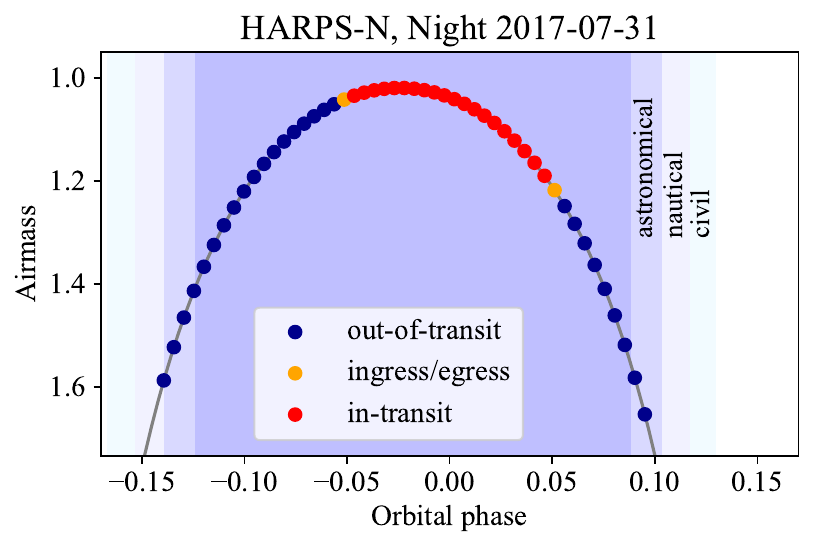}
    }
    \hfill
    \subfloat{
        \includegraphics[width=0.49\textwidth]{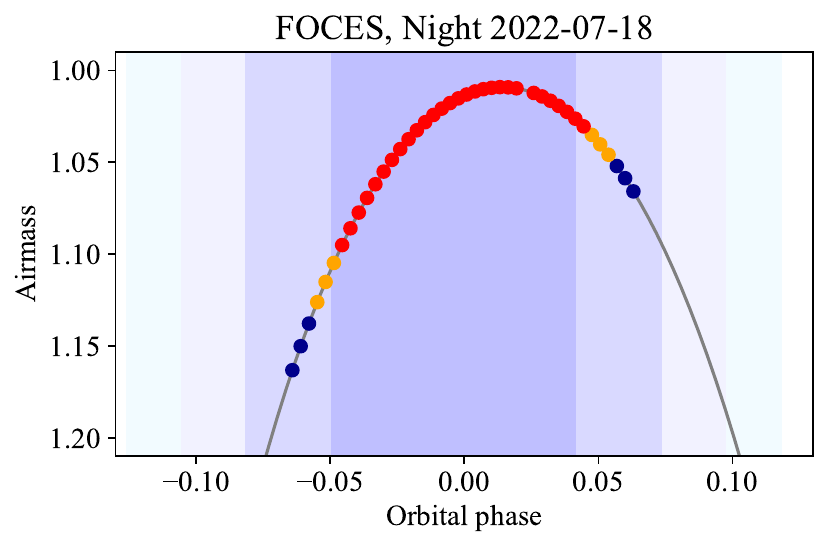}
    }


    \subfloat{
        \includegraphics[width=0.49\textwidth]{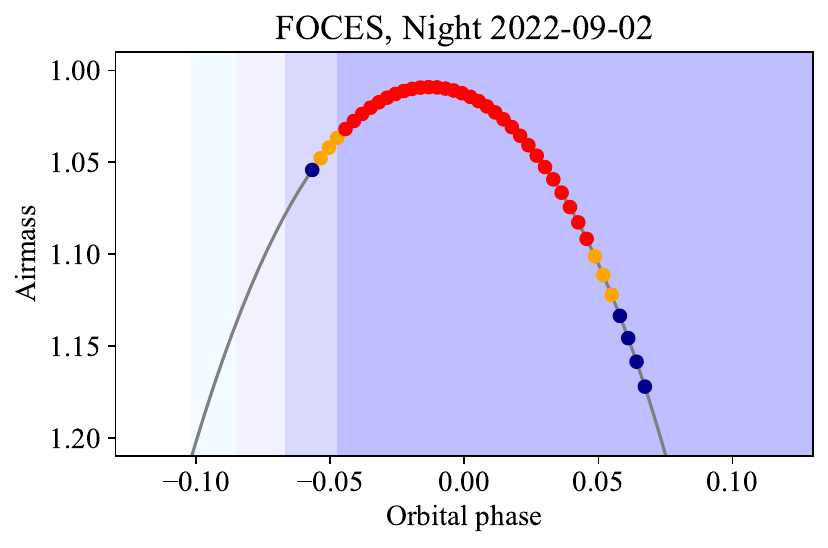}
    }
    \hfill
    \subfloat{
        \includegraphics[width=0.49\textwidth]{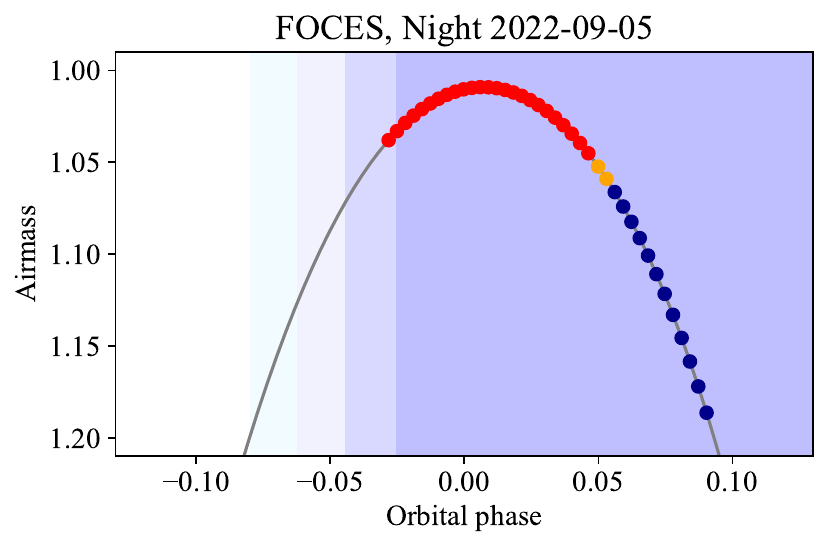}
    }

    \caption{Airmass plots for KELT-9\,b observations. Each data point represents an individual exposure frame. In-transit, ingress/egress, and out-of-transit exposures are indicated with red, orange, and blue. The civil, nautical, and astronomical twilight phases are indicated by the blue shaded regions. All transits were observed at low airmass. Parts of the transits observed with FOCES occurred during astronomical twilight given the short duration of the nights in summer.}
    \label{fig:airmassplots}
\end{figure*}

\section{KELT\text{-}9\,b atmospheric model}
\begin{figure*}[h]
    \centering
    \includegraphics[width=\textwidth]{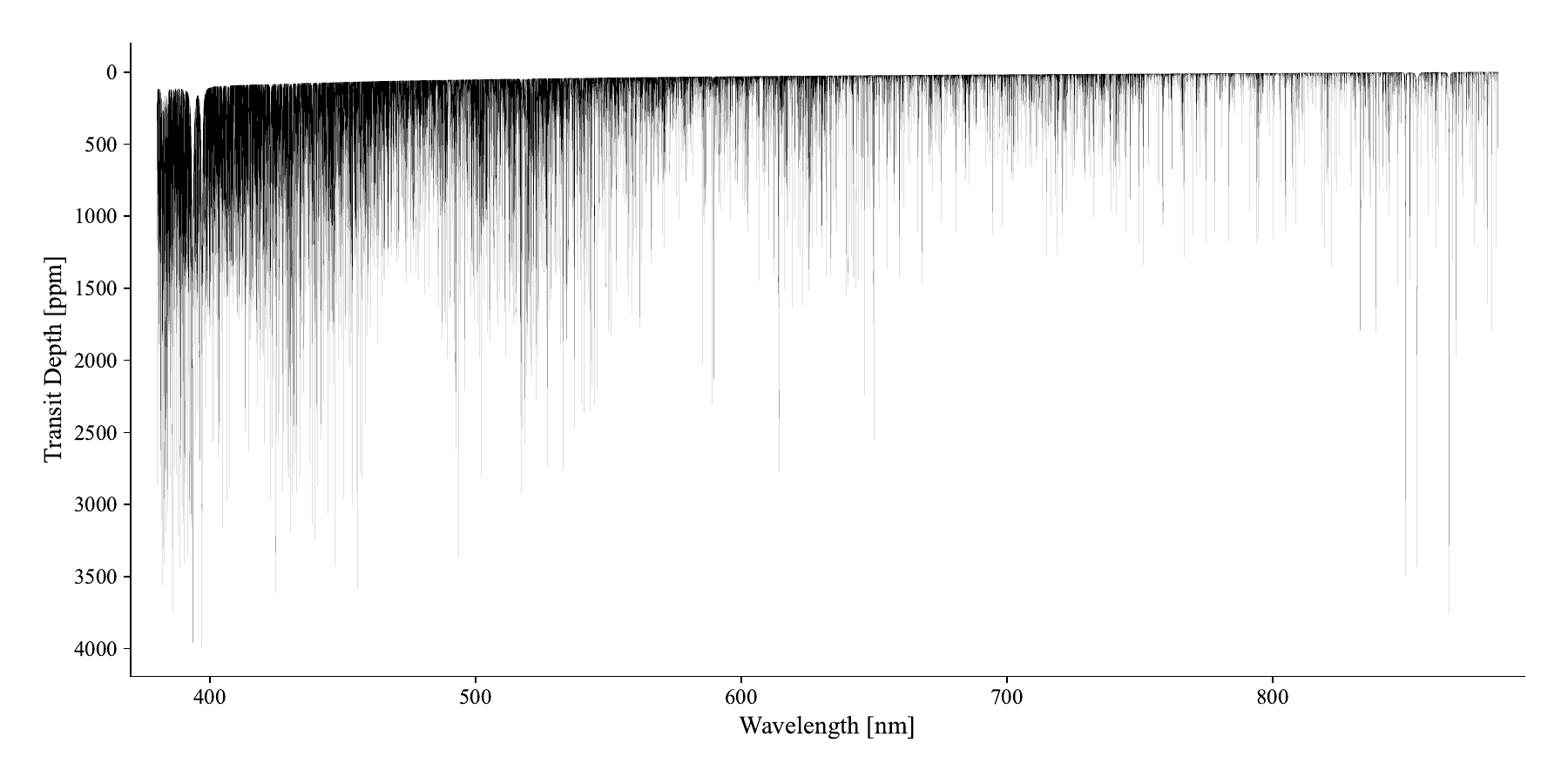}
    \caption{Radiative transfer model employed for model injection in this research. The model spans the wavelength range of FOCES. Line depth varies, typically on the order of a few 1000\,ppm.}
    \label{fig:model_for_injection}
\end{figure*}

In this section, we present the atmospheric model utilised for the injection procedure of this study, shown in Fig. \ref{fig:model_for_injection}. The model, on average, exhibits a relative transit depth on the order of a few tenths of a percent for the deepest lines. The high transit depth highlights why this planet is so conducive to atmospheric detections, which is a critical factor in WENDELSTEIN's success in detecting the planet's atmosphere in a single transit.

\section{Targets observable for Wendelstein}
In this section, we present three plots that illustrate the temperature dependence on line depth and its subsequent impact on detection significance. The first plot displays transmission templates from the Mantis network library \citep{Kitzmann_2021}, which were utilised in the emulator developed in this study. The remaining two plots depict the increase in detection significance for the 2000\,K and 4000\,K templates with a varying number of transits, illustrating the temperature-dependent detectability of specific species in exoplanet atmospheres. Given the aperture size of 2m class telescopes, it is anticipated that multiple transit observations will be necessary to discern the atmospheres of targets dimmer than 7.5 magnitude; nevertheless, achieving this objective remains a viable endeavour. Furthermore, we also include a comparison of the increase in S/N for 6.0 and 11.5 magnitude stars, which depicts how well the stacking results of the CCF align with Poisson trends.

\begin{figure*}[h]
    \centering
    \includegraphics[width=\textwidth]{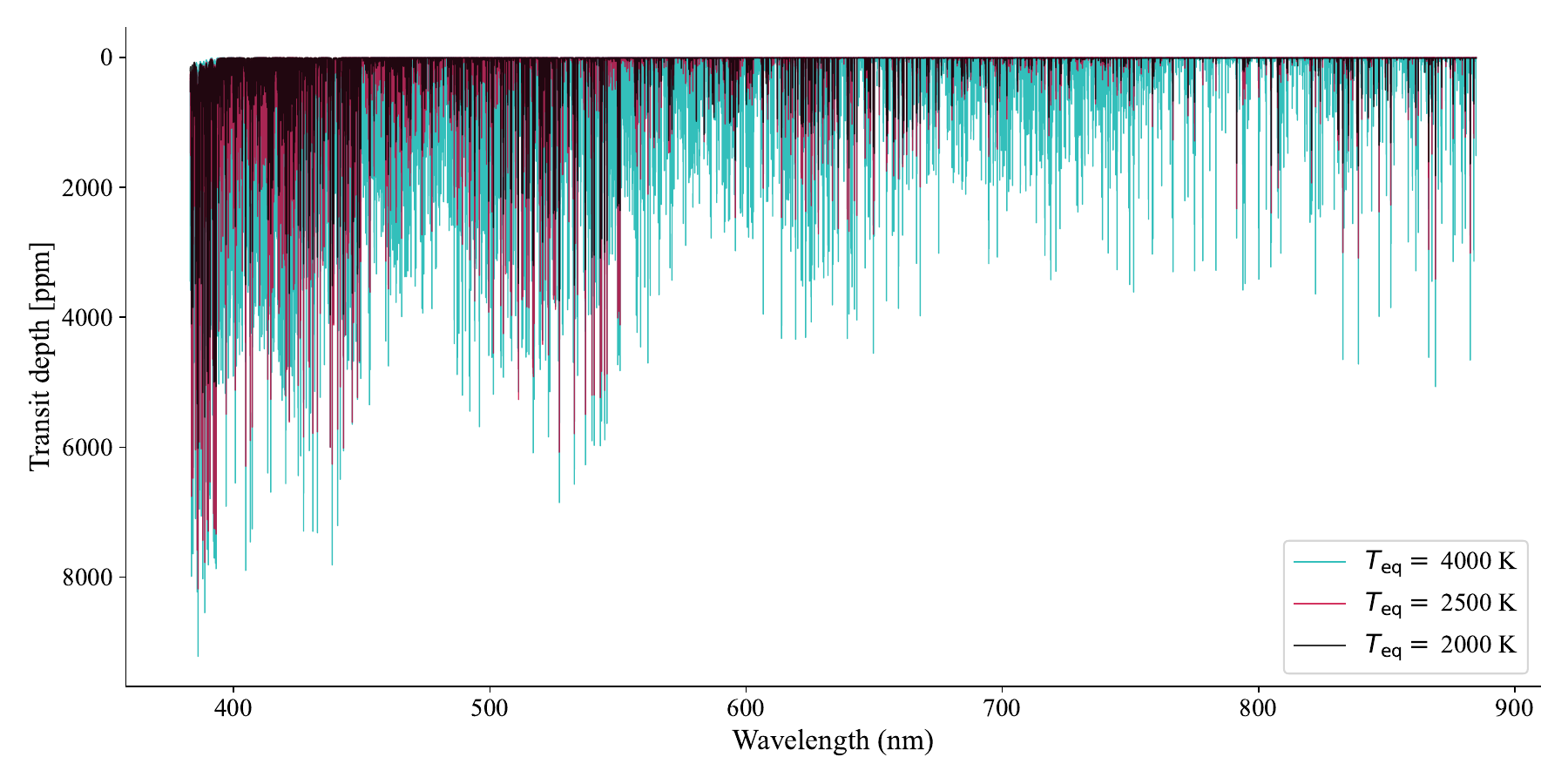}
    \caption{Variation in Fe I transmission with temperature. The plot illustrates three Fe I transmission templates used to emulate the transits. The 4000K template is represented in blue, the 2500\,K template in red, and the 2000\,K template in black. A discernible trend of increasing transmission depths is observed with escalating temperature, wherein more lines become pronounced and stronger. This enhancement in line strength with temperature increment effectively introduces more structure to the auto-correlation function, elucidating the temperature-dependent behaviour of Fe I transmission in the emulation of transits.}
    \label{fig:equib_templates_comparison}
\end{figure*}

\begin{figure*}[h]
    \centering
    \includegraphics[width=\textwidth]{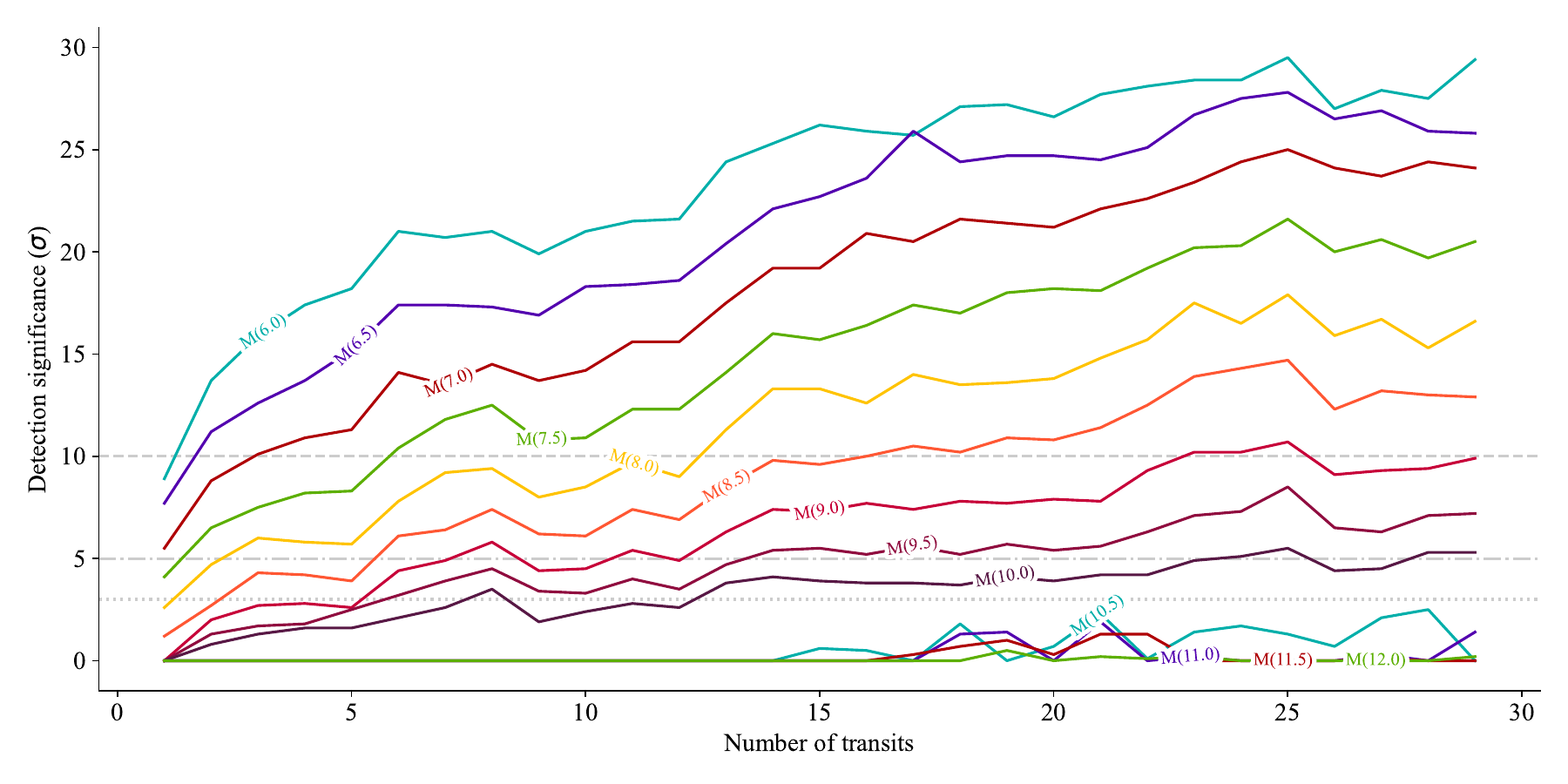}
    \caption{Detection significance trends for the 2000\,K following the same process as Fig. \ref{fig:transit_significances_trend_2500}.}
    \label{fig:transit_significances_trend_2000}
\end{figure*}

\begin{figure*}[h]
    \centering
    \includegraphics[width=\textwidth]{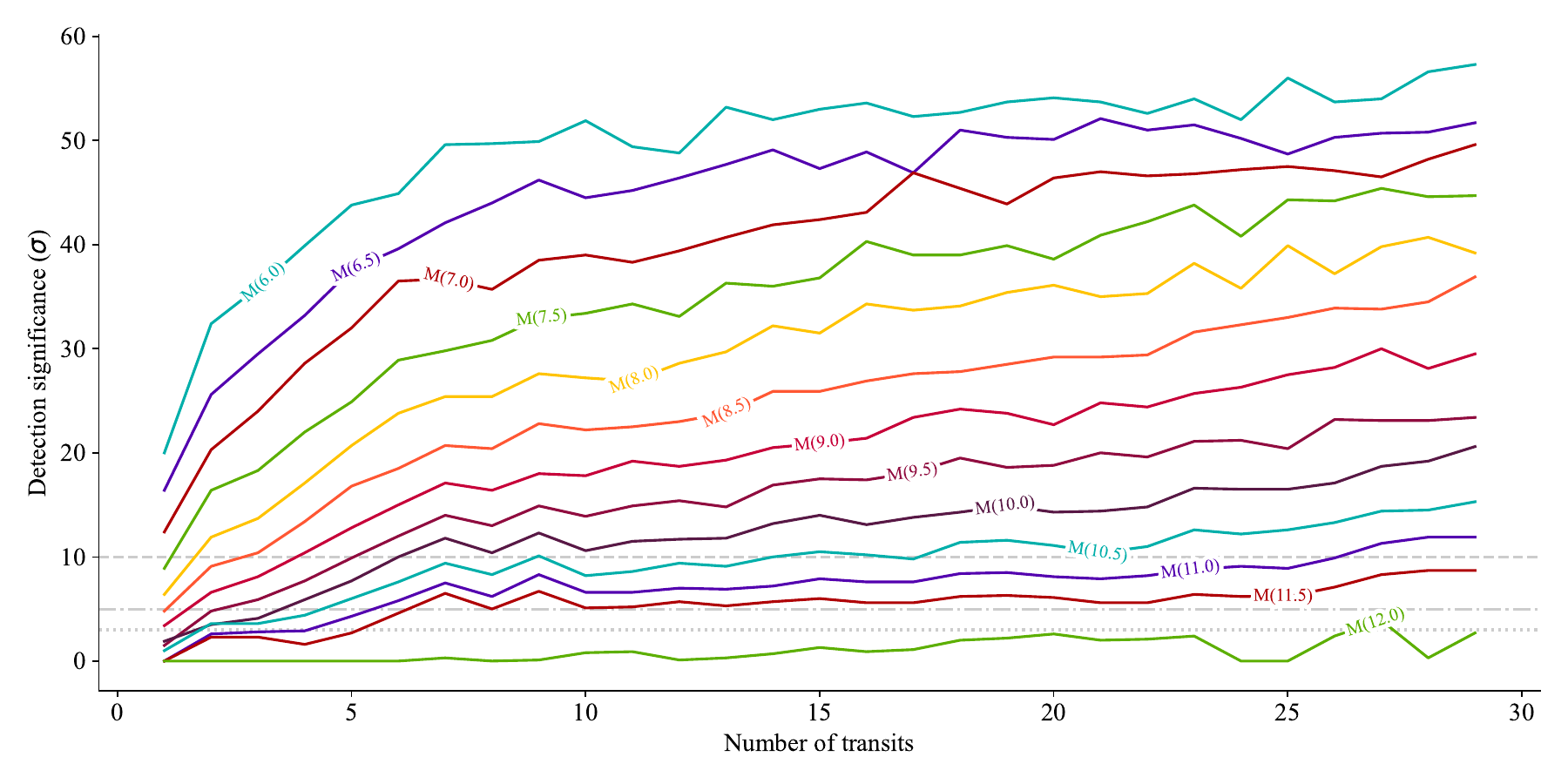}
    \caption{Detection significance trends for the 4000\,K following the same process as Fig. \ref{fig:transit_significances_trend_2500}.}
    \label{fig:transit_significances_trend_4000}
\end{figure*}

\begin{figure*}[h]
    \centering
    \includegraphics[width=\textwidth]{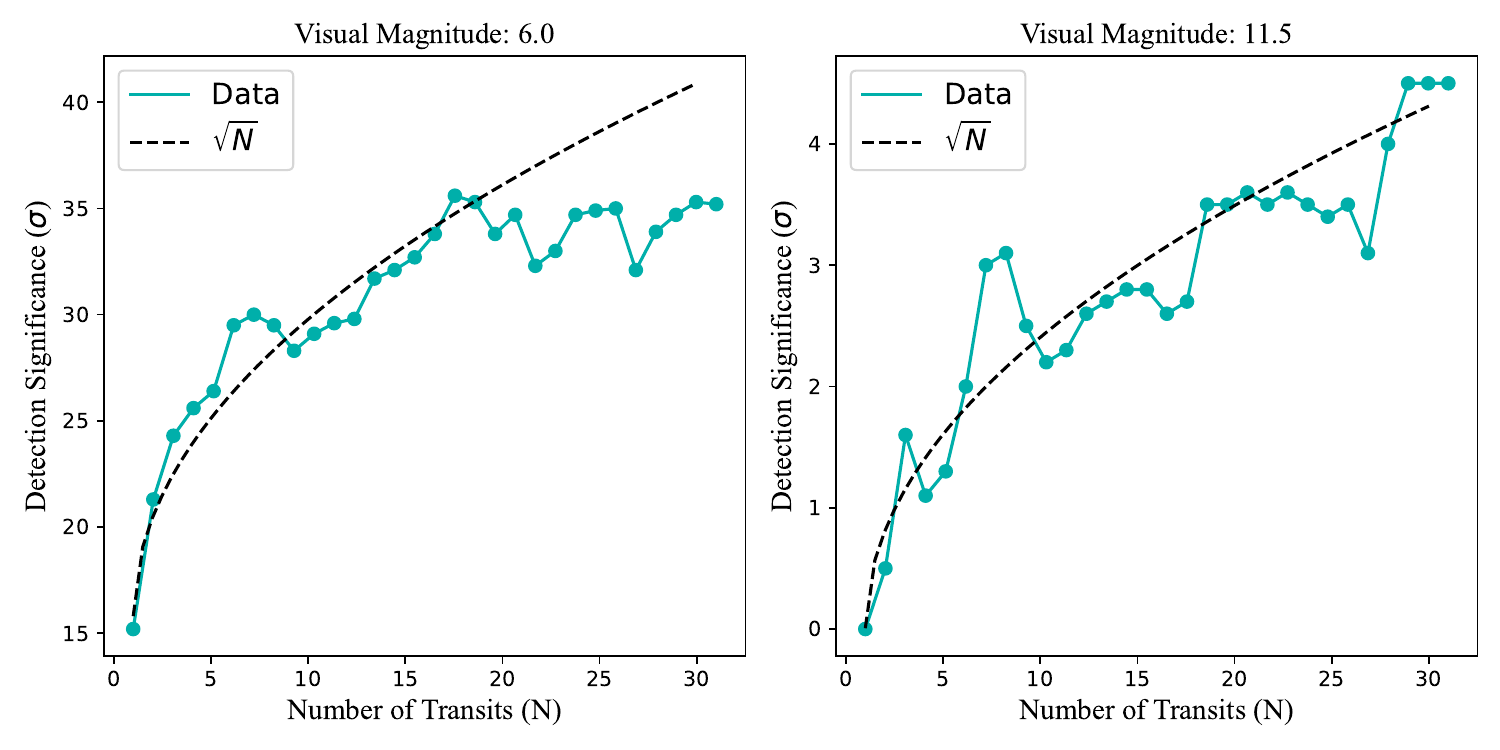}
    \caption{Comparison of the predicted Poisson trend with actual significance in stacked transits, with the transit emulator. Each plot features the number of transits against the detection significance, depicted in teal, alongside a dashed line representing the square root fit of the first ten stacked transits. The left panel illustrates results from stacking transits around a star of magnitude 6.0. The right panel shows a similar trend for a star of magnitude 11.5.}
    \label{fig:divergance_from_poisson}
\end{figure*}

\end{document}